\documentclass[english]{revtex4-1}
\usepackage[T1]{fontenc}
\usepackage[latin9]{inputenc}
\usepackage{array}
\usepackage{float}
\usepackage{multirow}
\usepackage{amsmath}
\usepackage{graphicx}

\makeatletter

\providecommand{\tabularnewline}{\\}

\makeatother

\usepackage{babel}
\begin{document}

\title{Possible discovery channel for fourth chiral family up-quark at the
LHC}

\author{S. Beser}
\email{sbeser@etu.edu.tr}

\affiliation{TOBB University of Economics and Technology, Ankara, Turkey }

\author{U. Kaya}
\email{ukaya@etu.edu.tr}

\affiliation{TOBB University of Economics and Technology, Ankara, Turkey }

\affiliation{Ankara University, Ankara, Turkey}

\author{B. B. Oner}
\email{b.oner@etu.edu.tr}

\affiliation{TOBB University of Economics and Technology, Ankara, Turkey }

\author{S. Sultansoy}
\email{ssultansoy@etu.edu.tr}

\affiliation{TOBB University of Economics and Technology, Ankara, Turkey }

\affiliation{ANAS Institute of Physics, Baku, Azerbaijan}
\begin{abstract}
Resonant production of fourth chiral family up quark at the LHC via
anomalous interactions have been analyzed. It is shown that search
for resonances in $W^{+}b$ final states could lead to discovery of
the fourth chiral family and simultaneously determine scale of the
new physics, presumabely related to the quark and lepton compositeness.
Obtained results emphasize an importance of W-leading jet invariant
mass analysis in search for W+jets final states at the LHC, both with
and without b-tagging.
\end{abstract}
\maketitle

\section{Introduction}

It is known that the Standard Model does not fix the number of fermion
families. This number should be less than 9 in order to preserve asymptotic
freedom and more than 2 in order to provide CP violation. According
to the LEP data on Z decays, number of chiral families with light
neutrinos ($M_{\nu}\ll M_{Z}$) is equal to 3, whereas extra families
with heavy neutrinos are not forbidden. The fourth chiral family was
widely discussed thirty years ago (see, for example \cite{key-1,key-2}).
However, the topic was pushed off the agenda due to the misinterpretation
of the LEP data. 

Twenty years later 3 workshops on the fourth SM family \cite{key-3,key-4,key-5}
were held (for summary of the first and third workshops see \cite{key-6}
and \cite{key-7}, respectively). Main motivation was Flavor Democracy
\cite{key-8,key-9,key-10} which naturally provides heavy fourth family
fermions including neutrino (consequences of Flavor Democracy Hypothesis
for different models, including MSSM and $E_{6}$, have been considered
in \cite{key-11,key-12}). In addition, fourth family gives opportunity
to explain baryon asymmetry of Universe, it can accommodate emerging
possible hints of new physics in rare decays of heavy mesons etc (see
\cite{key-6} and references therein). Phenomenological papers on
direct production (including anomalous resonant production) of the
SM4 fermions at different colliders are reviewed in \cite{key-13}
(see tables VI and VII in \cite{key-13}). 

This activity has almost ended due to misinterpretation of the LHC
data on the Higgs decays. It should be emphasized that these data
exclude the minimal SM4 with one Higgs doublet, whereas non-minimal
SM4 with extended Higgs sector are still allowed \cite{key-14,key-15}.
On the other hand, partial wave unitarity puts an upper limit around
700 GeV on the masses of fourth SM family quarks \cite{key-16}, which
is almost excluded by the recent ATLAS and CMS data on search for
pair production. For example, ATLAS $\sqrt{s}=8$ TeV data with 20.3
$fb^{-1}$ integrated luminosity excludes new chiral quarks with mass
below 690 GeV at 95\% confidence level assuming BR($Q\rightarrow Wq$)=1
\cite{key-17}. 

Even if SM4 may be excluded by the LHC soon, this is not the case
for the general chiral fourth family (C4F). Therefore, ATLAS and CMS
should continue a search for C4F up to kinematical limits. Concerning
pair production, rescaling of the ATLAS lower bound using collider
reach framework \cite{key-18} shows that LHC with $\sqrt{s}=13$
TeV will give opportunity to cover $M_{u_{4}}$ up to 0.94, 1.25,
1.50, and 2.13 TeV with integrated luminosities 20, 100, 300 and 3000
$fb^{-1}$, respectively. 

This study is motivated by a recent paper \cite{key-19}, which shows
that in order to interprete 125 GeV scalar boson as dilaton the mass
of the fourth family quarks should exceed 2-2.5 TeV; therefore fourth
family quarks can not be observed at the LHC via pair production.
For this reason, we consider possible resonant production of C4F quarks
through their anomalous interactions with light quarks. In Section
II standard and anomalous decays of the C4F quarks are considered.
Resonant production of fourth chiral family quarks is analyzed in
Section III. Finally, conclusions and recommendations are given in
Section IV. 

\section{Standard and anomalous decays of the C4F quarks}

The effective Lagrangian for anomalous magnetic type interactions
of the fourth family quarks is given as \cite{key-20,key-21,key-22}:

\begin{equation}
L=\underset{q_{i}}{\sum}\frac{\kappa_{\gamma}^{q_{i}}}{\varLambda}e_{q}g_{e}\bar{q}_{4}\sigma_{\mu\nu}q_{i}F^{\mu\nu}+\underset{q_{i}}{\sum}\frac{\kappa_{Z}^{q_{i}}}{2\varLambda}g_{Z}\bar{q}_{4}\sigma_{\mu\nu}q_{i}Z^{\mu\nu}+\underset{q_{i}}{\sum}\frac{\kappa_{g}^{q_{i}}}{\varLambda}g_{s}\bar{q}_{4}\sigma_{\mu\nu}T^{a}q_{i}G_{a}^{\mu\nu}+H.c.\label{eq:1}
\end{equation}

\noindent where $F^{\mu\nu}$, $Z^{\mu\nu}$, and $G^{\mu\nu}$ are
the field strength tensors of the gauge bosons, $\sigma_{\mu\nu}$
is the antisymmetric tensor, $T^{a}$ are Gell-Mann matrices, $e_{q}$
is electric charge of quark, $g_{e}$, $g_{Z}$ and $g_{s}$ are electromagnetic,
neutral weak, and strong coupling constants, respectively. $g_{Z}=g_{e}/cos\theta_{W}$,
where $\theta_{W}$is the Weinberg angle, $\kappa_{\gamma}$, $\kappa_{Z}$,
and $\kappa_{g}$ are the strength of anomalous couplings with photon,
Z boson and gluon, respectively. $\Lambda$ is the cutoff scale for
new physics. For numerical calculations, we implement Lagrangian (\ref{eq:1})
into CalcHEP package \cite{key-23}.

The partial decay widths of $u_{4}$ for SM ($u_{4}\rightarrow W^{+}q$,
where $q=d,s,b$) and anomalous ($u_{4}\rightarrow\gamma q$, $u_{4}\rightarrow Zq$,
$u_{4}\rightarrow gq$, where $q=u,c,t$) modes are given below \cite{key-13}:

\begin{equation}
\Gamma(u_{4}\rightarrow W^{+}q)=\frac{|V_{u_{4}q}|^{2}\alpha_{e}m_{u_{4}}^{3}}{16m_{W}^{2}sin^{2}\theta_{W}}\varsigma_{W}\sqrt{\varsigma_{0}},
\end{equation}

\noindent where $\varsigma_{W}=(1+x_{q}^{4}+x_{q}^{2}x_{W}^{2}-2x_{q}^{2}-2x_{W}^{4}+x_{W}^{2})$,
$\varsigma_{0}=(1+x_{W}^{4}+x_{q}^{4}-2x_{W}^{2}-2x_{q}^{2}-2x_{W}^{2}x_{q}^{2})$,
$x_{q}=(m_{q}/m_{u_{4}})$, and $x_{W}=(m_{W}/m_{u_{4}})$,

\begin{equation}
\Gamma(u_{4}\rightarrow Zq)=\frac{\alpha_{e}m_{u_{4}}^{3}}{16cos^{2}\theta_{W}sin^{2}\theta_{W}}\left(\frac{\kappa_{Z}^{q}}{\Lambda}\right)^{2}\varsigma_{Z}\sqrt{\varsigma_{1}},
\end{equation}

\noindent where $\varsigma_{Z}=(2-x_{Z}^{4}+x_{Z}^{2}-4x_{q}^{2}-x_{q}^{2}x_{Z}^{2}-6x_{q}x_{Z}^{2}+2x_{q}^{4})$,
$\varsigma_{1}=(1+x_{Z}^{4}+x_{q}^{2}-2x_{Z}^{2}-2x_{q}^{2}-2x_{Z}^{2}x_{q}^{2})$,
and $x_{Z}=(m_{Z}/m_{u_{4}})$,

\begin{equation}
\Gamma(u_{4}\rightarrow gq)=\frac{2\alpha_{s}m_{u_{4}}^{3}}{3}\left(\frac{\kappa_{g}^{q}}{\Lambda}\right)^{2}\varsigma_{2},
\end{equation}

\noindent where $\varsigma_{2}=(1-3x_{q}^{2}+3x_{q}^{4}-x_{q}^{6})$,

\begin{equation}
\Gamma(u_{4}\rightarrow\gamma q)=\frac{\alpha_{e}m_{u_{4}}^{3}Q_{q}^{2}}{2}\left(\frac{\kappa_{\gamma}^{q}}{\Lambda}\right)^{2}\varsigma_{2},
\end{equation}

The partial decay widths of $d_{4}$ for SM ($d_{4}\rightarrow W^{-}q$,
where $q=u,c,t$) and anomalous ($d_{4}\rightarrow\gamma q$, $d_{4}\rightarrow Zq$,
$d_{4}\rightarrow gq$, where $q=d,s,b$) modes are given below:

\begin{equation}
\Gamma(d_{4}\rightarrow W^{-}q)=\frac{|V_{qd_{4}}|^{2}\alpha_{e}m_{d_{4}}^{3}}{16m_{W}^{2}sin^{2}\theta_{W}}\chi_{W}\sqrt{\chi_{0}},
\end{equation}

\noindent where $\chi_{W}=(1+y_{q}^{4}+y_{q}^{2}y_{W}^{2}-2x_{q}^{2}-2y_{W}^{4}+y_{W}^{2})$,
$\chi_{0}=(1+y_{W}^{4}+y_{q}^{4}-2y_{W}^{2}-2y_{q}^{2}-2y_{W}^{2}x_{q}^{2})$,
$y_{q}=(m_{q}/m_{d_{4}})$, and $y_{W}=(m_{W}/m_{d_{4}})$,

\begin{equation}
\Gamma(d_{4}\rightarrow Zq)=\frac{\alpha_{e}m_{d_{4}}^{3}}{16cos^{2}\theta_{W}sin^{2}\theta_{W}}\left(\frac{\kappa_{Z}^{q}}{\Lambda}\right)^{2}\chi_{Z}\sqrt{\chi_{1}},
\end{equation}

\noindent where $\chi_{Z}=(2-y_{Z}^{4}+y_{Z}^{2}-4y_{q}^{2}-y_{q}^{2}y_{Z}^{2}-6y_{q}y_{Z}^{2}+2y_{q}^{4})$,
$\chi_{1}=(1+y_{Z}^{4}+y_{q}^{2}-2y_{Z}^{2}-2y_{q}^{2}-2y_{Z}^{2}y_{q}^{2})$,
and $y_{Z}=(m_{Z}/m_{d_{4}})$,

\begin{equation}
\Gamma(d_{4}\rightarrow gq)=\frac{2\alpha_{s}m_{d_{4}}^{3}}{3}\left(\frac{\kappa_{g}^{q}}{\Lambda}\right)^{2}\chi_{2},
\end{equation}

\noindent where $\chi_{2}=(1-3y_{q}^{2}+3y_{q}^{4}-y_{q}^{6})$,

\begin{equation}
\Gamma(d_{4}\rightarrow\gamma q)=\frac{\alpha_{e}m_{d_{4}}^{3}Q_{q}^{2}}{2}\left(\frac{\kappa_{\gamma}^{q}}{\Lambda}\right)^{2}\chi_{2}.
\end{equation}

Hereafter, we assume the dominance of CKM mixings between the fourth
and third families for standard decays and dominance of anomalous
interactions between fourth and first families. Partial decay widths
of $u_{4}$ to $W^{+}b$ (assuming $V_{u_{4}b}=0.1$) and $ug$ (assuming
$\kappa=1$ and $\Lambda=100$ TeV) channels are presented in Figures
1 and 2, respectively. Branching ratios for $u_{4}\rightarrow W^{+}b$
are presented in Figures 3 and 4. It is seen that this decay channel
is dominant in large intervals of $\Lambda$ and $V_{u_{4}b}$. 

\begin{figure}[H]
\begin{centering}
\includegraphics[scale=0.12]{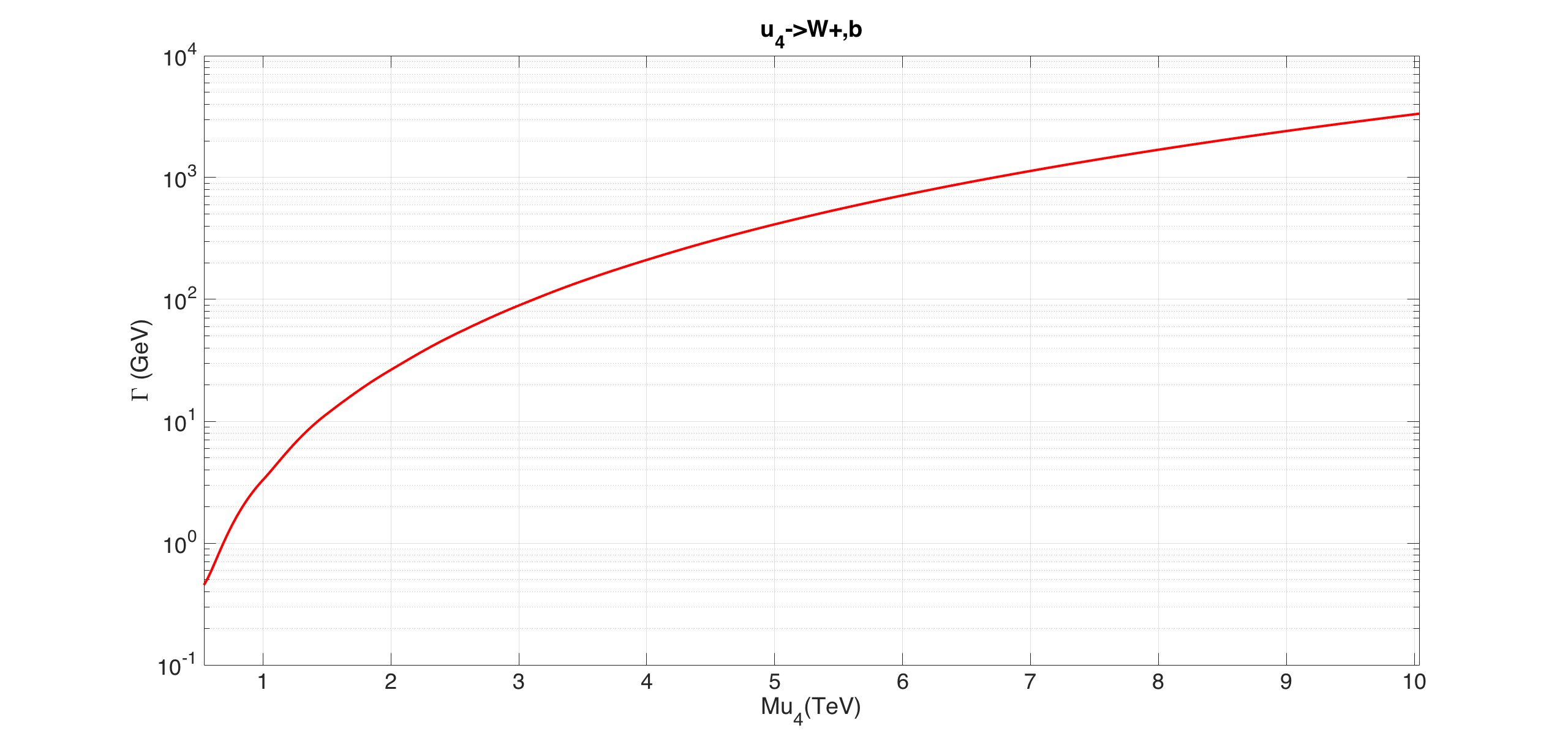}
\par\end{centering}
\caption{$\Gamma(u_{4}\rightarrow W^{+}b)$ with $Vu_{4}b=0.1$.}
\end{figure}

\begin{figure}[H]
\begin{centering}
\includegraphics[scale=0.12]{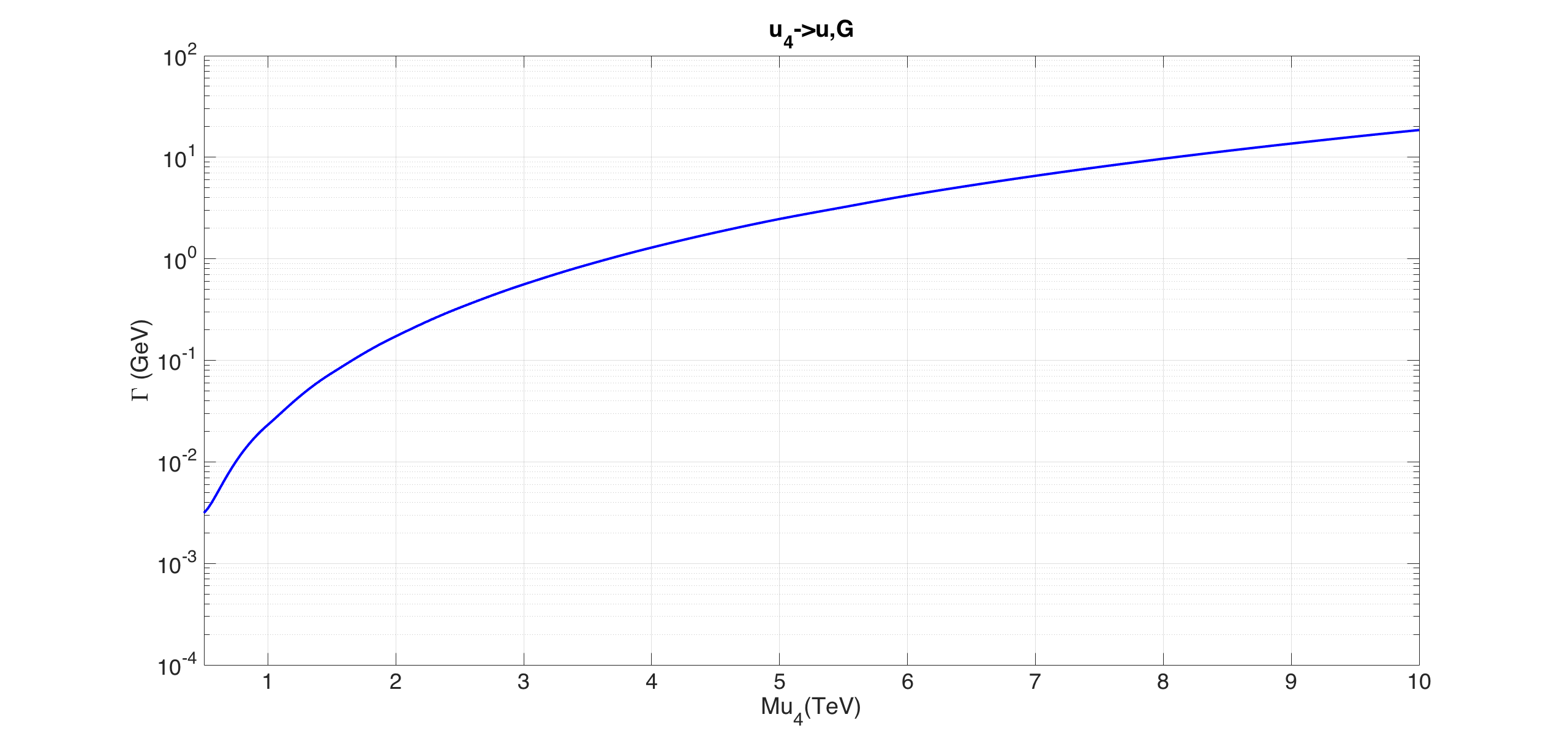}
\par\end{centering}
\caption{$\Gamma(u_{4}\rightarrow ug)$ with $\kappa_{g}^{u}=1$ and $\Lambda=100$
TeV.}
\end{figure}

\begin{figure}[H]
\begin{centering}
\includegraphics[scale=0.12]{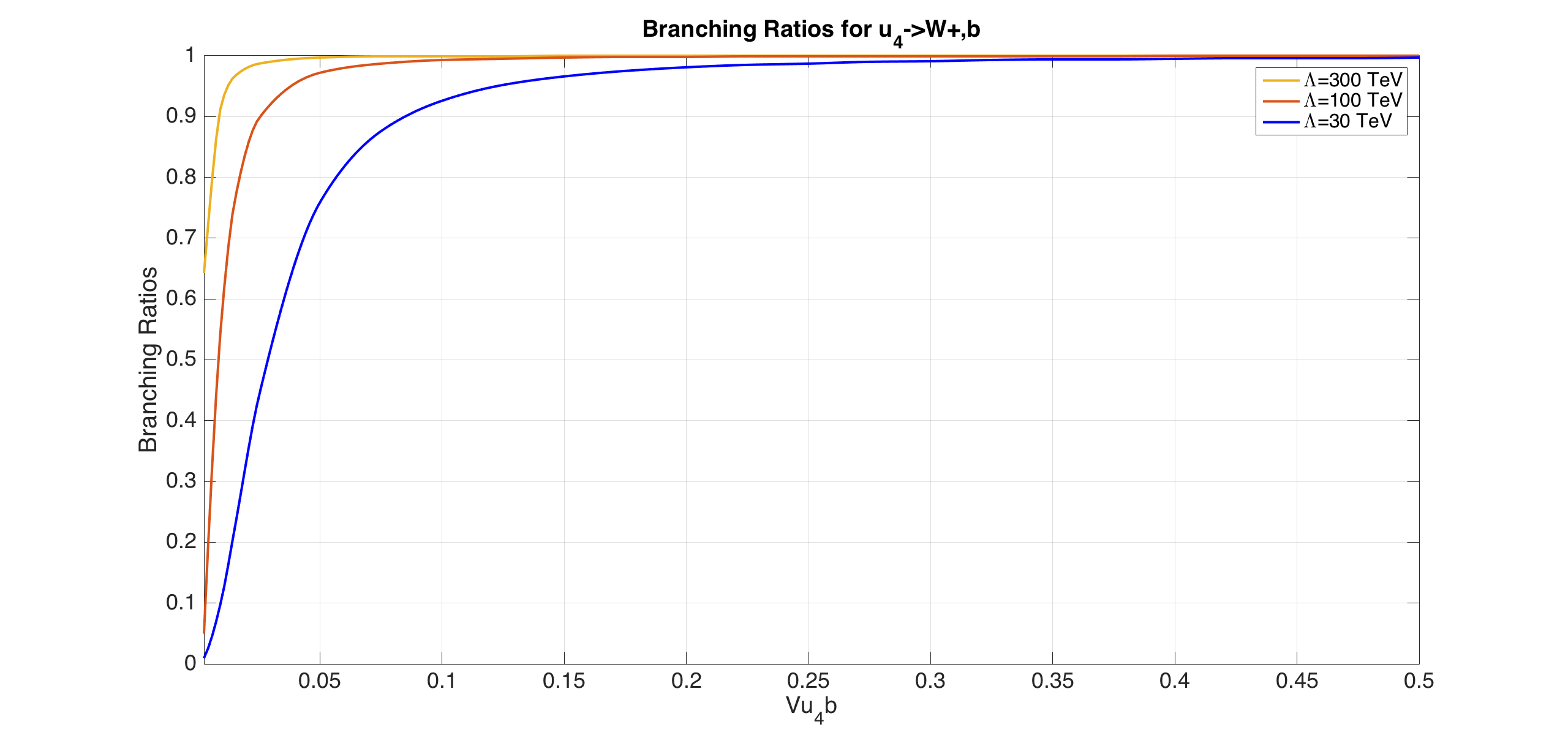}
\par\end{centering}
\caption{$BR(u_{4}\rightarrow W^{+}b)$ for $M_{u_{4}}=3$ TeV and $\Lambda=30,\,100,\,300$
TeV.}
\end{figure}

\begin{figure}[H]
\begin{centering}
\includegraphics[scale=0.12]{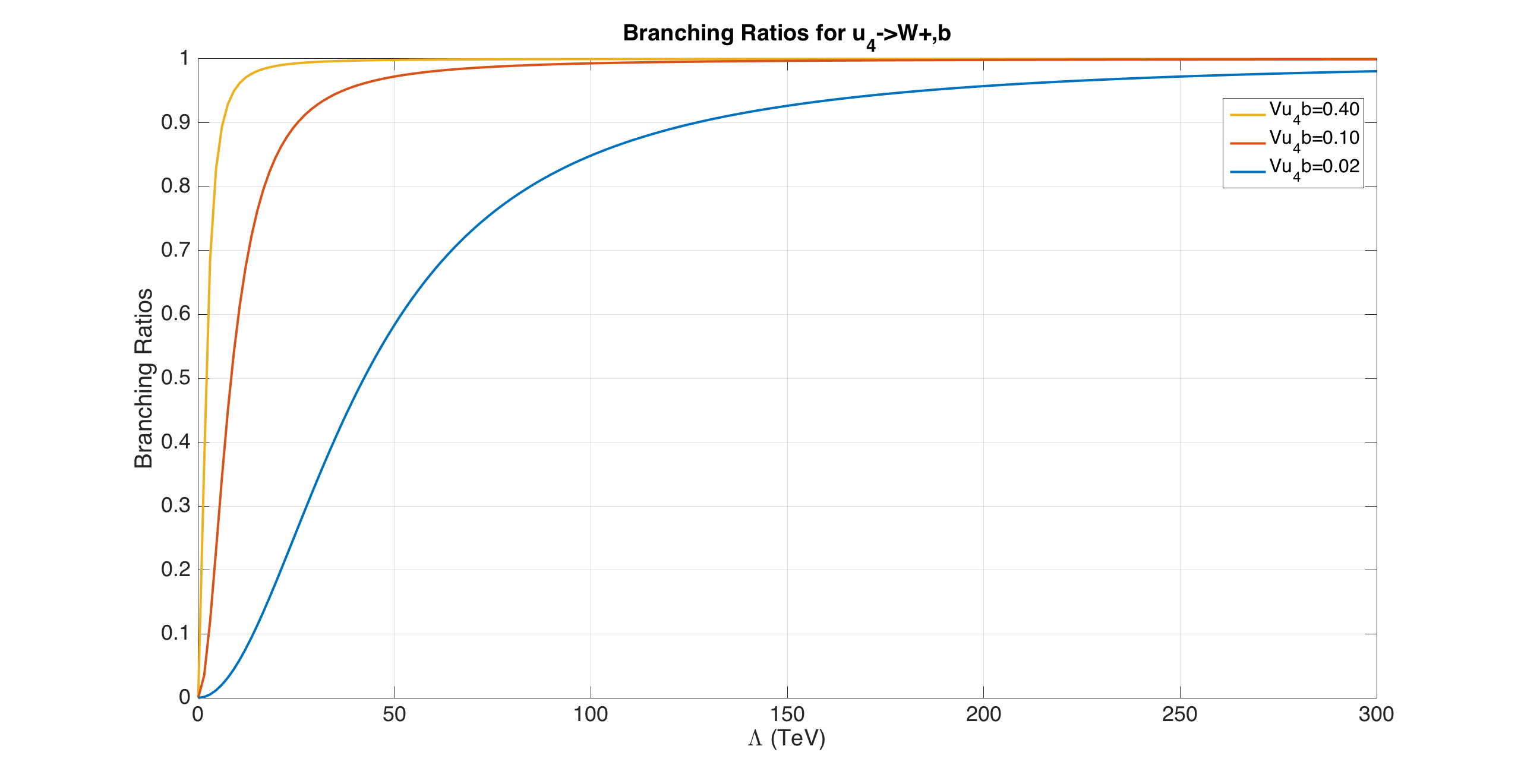}
\par\end{centering}
\caption{$BR(u_{4}\rightarrow W^{+}b)$ for $M_{u_{4}}=3$ TeV and $Vu_{4}b=0.4,\,0.1,\,0.02$.}
\end{figure}

\section{Resonant Product\i on of $u_{4}$ quarks at the lhc}

Anomalous resonant production of $u_{4}$ quarks with dominant $u_{4}\rightarrow W^{+}b$
decays could provide unique opportunity for the discovery of the fourth
chiral family at the LHC because this signature can not be imitated
by other BSM particles. Corresponding Feynman diagram is given in
Figure 5. In Figure 6 we present dependence of resonance $u_{4}$
production cross section at the LHC with $\sqrt{s}=13$ TeV for different
values of $\Lambda$. For numerical calculations, CalcHEP with CTEQ6l
pdf has been used {[}24{]}. 

Following study is performed for leptonic decay of W boson, namely
$W^{+}\rightarrow e^{+}\nu_{e}$. In order to determine discovery
cuts, $p_{T}$ distributions of b-quarks, positrons and neutrinos
(missing transverse momentum) for signal and background proccesses
have been analyzed. It was seen that $p_{T}^{b}>200$ GeV and $p_{T}^{e^{+}},\,p_{T}^{miss}>100$
GeV drastically reduce background whereas signal is almost unchanged.
In addition we choose $0.9M_{u_{4}}<M_{inv}<1.1M_{u_{4}}$ for invariant
mass window. Statistical significance is deterimined as:

\begin{equation}
SS=\sqrt{2[(S+B)\,ln(1+(S/B))-S]}
\end{equation}

\noindent where $S$ and $B$ are number of events of signal and background,
respectively. With these cuts, for $\varLambda=100$ TeV we obtain
discovery limits on the mass of $u_{4}$ as 3.5, 4.3 and 5.9 TeV for
$L_{int}=$ 100, 300 and 3000 $fb^{-1}$, respectively. Discovery
of $u_{4}$ at the LHC via the channel under the consideration will
simultaneously determine $\Lambda$ scale. Achievable $\Lambda$ values
for different $M_{u_{4}}$ and $L_{int}$ are presented in Table I.

\begin{table}[b]
\caption{Achievable $\Lambda$ values for different $M_{u_{4}}$ and $L_{int}$.}

\centering{}%
\begin{tabular}{|c|c|c|c|}
\hline 
$M_{u_{4}}$, TeV & $L_{int}$, $fb^{-1}$ & $\Lambda$, TeV ($SS=3$) & $\Lambda$, TeV ($SS=5$)\tabularnewline
\hline 
\hline 
\multirow{3}{*}{1} & 3000 & 680 & 540\tabularnewline
\cline{2-4} 
 & 300 & 380 & 292\tabularnewline
\cline{2-4} 
 & 100 & 286 & 220\tabularnewline
\hline 
\multirow{3}{*}{2} & 3000 & 510 & 395\tabularnewline
\cline{2-4} 
 & 300 & 285 & 219\tabularnewline
\cline{2-4} 
 & 100 & 218 & 163\tabularnewline
\hline 
\multirow{3}{*}{3} & 3000 & 420 & 324\tabularnewline
\cline{2-4} 
 & 300 & 231 & 175\tabularnewline
\cline{2-4} 
 & 100 & 171 & 127\tabularnewline
\hline 
\multirow{3}{*}{4} & 3000 & 285 & 218\tabularnewline
\cline{2-4} 
 & 300 & 151 & 109\tabularnewline
\cline{2-4} 
 & 100 & 109 & 76\tabularnewline
\hline 
\multirow{3}{*}{5} & 3000 & 208 & 167\tabularnewline
\cline{2-4} 
 & 300 & 103 & 70\tabularnewline
\cline{2-4} 
 & 100 & 68 & 38\tabularnewline
\hline 
\multirow{3}{*}{5.5} & 3000 & 168 & 122\tabularnewline
\cline{2-4} 
 & 300 & 78 & 46\tabularnewline
\cline{2-4} 
 & 100 & 45 & -\tabularnewline
\hline 
\multirow{3}{*}{6} & 3000 & 137 & 96\tabularnewline
\cline{2-4} 
 & 300 & 11 & 22\tabularnewline
\cline{2-4} 
 & 100 & - & -\tabularnewline
\hline 
\end{tabular}
\end{table}

\begin{figure}[t]
\begin{centering}
\includegraphics[scale=0.75]{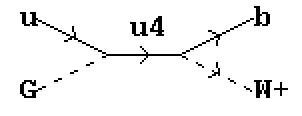}
\par\end{centering}
\caption{Feynman diagram for resonant $u_{4}$ prodution at the LHC.}
\end{figure}

\begin{figure}[H]
\begin{centering}
\includegraphics[scale=0.15]{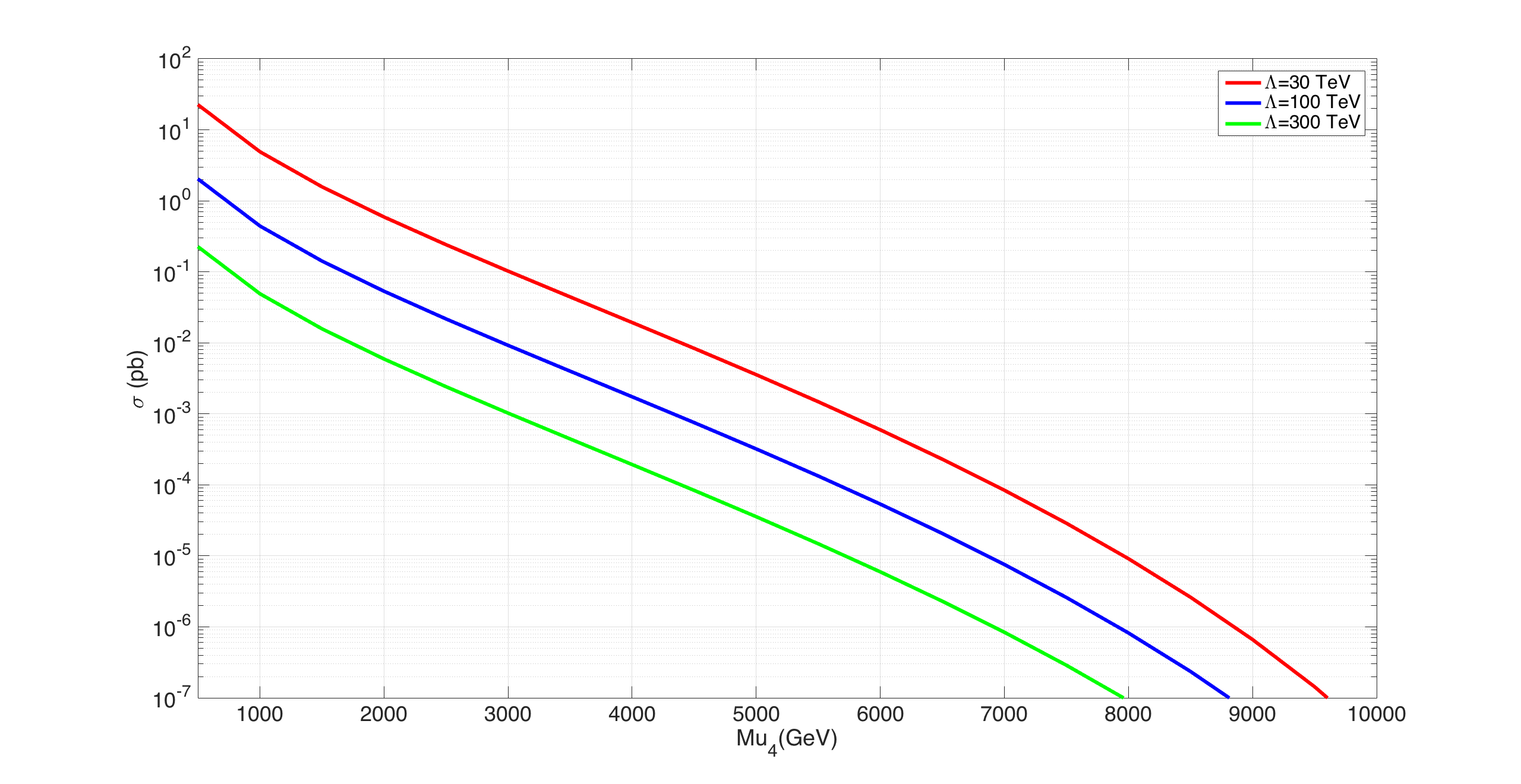}
\par\end{centering}
\caption{Cross section of resonant $u_{4}$ production at the LHC with $\sqrt{s}=13$
TeV for different $\Lambda$ values.}
\end{figure}

\section{CONCLUSION}

It is seen that search for resonance in $W^{+}b$ final state at the
LHC could lead to discovery of the fourth chiral family and simultaneously
determine scale of the new physics, presumably related to the quark
and lepton compositeness. In this paper only special case, namely
$ug\rightarrow u_{4}\rightarrow W^{+}b$ has been analyzed ($W^{-}j$
final states are suppressed since they are originated from sea $\bar{u}$
quarks). In the same manner $d_{4}$ quarks can also be produced with
cross section approximately half of $u_{4}$. In this case $d_{4}$
has following decay chain: $d_{4}\rightarrow W^{-}t\rightarrow W^{-}W^{+}b$. 

If CKM mixings with the second (first) family are dominant, in spite
of dominance of CKM mixings with the third family considered in this
study, main decay modes will be $u_{4}\rightarrow W^{+}s(W^{+}d)$
and $d_{4}\rightarrow W^{-}c(W^{-}u)$. In this case both $W^{+}j$
and $W^{-}j$ final states are expected with the ratio of 2:1. 

Fourth family quarks may also be produced via anomalous couplings
with second and third family quarks. In this case production cross
sections are suppressed since initial states includes sea quarks.
Anomalous coupling with second family quarks results in the processes
$sg\rightarrow d_{4}\rightarrow W^{-}b$, $\bar{s}g\rightarrow\bar{d}_{4}\rightarrow W^{+}\bar{b}$,
$cg\rightarrow u_{4}\rightarrow W^{+}b$, $\bar{c}g\rightarrow\bar{u}_{4}\rightarrow W^{-}\bar{b}$
(if CKM mixings with first two families are dominant b-jets are replaced
by lighter quark-jets). Anomalous coupling with third family quarks
contributes to the production of chiral fourth family down quarks
via $bg\rightarrow d_{4}\rightarrow W^{-}b$, $\bar{b}g\rightarrow\bar{d}_{4}\rightarrow W^{+}\bar{b}$.

Finally, it is possible that anomalous decay modes of fourth family
quarks may be dominant. Unfortunately, in this case, identification
of fourth family quarks will be problematic since, for example, excited
quarks (q{*}) has similar signatures.

In conclusion, results of this study emphasize an importance of W-leading
jet invariant mass analysis in search for W+jets final states at the
LHC, both with and without b-tagging.
\begin{acknowledgments}
Authors are grateful to Y. C. Acar, M. Sahin and G. Unel for useful
discussions. This work is partially supported by TUBITAK under the
grant no 114F337.
\end{acknowledgments}

\end{document}